\documentclass[fleqn,usenatbib]{mnras}

\usepackage{newtxtext,newtxmath}
\usepackage[T1]{fontenc}
\usepackage{ae,aecompl}
\usepackage{graphicx}
\usepackage{amsmath}
\usepackage[mathscr]{euscript}

\let\mathscr\relax 
\usepackage[scr]{rsfso}
\usepackage{enumitem}
\usepackage{tabularx, booktabs, makecell}
\usepackage{siunitx}
\newcommand{\daa}{\Delta\alpha/\alpha}

\DeclareMathAlphabet{\mathpzc}{OT1}{pzc}{m}{it}

\title[Model non-uniqueness and varying $\alpha$]{Non-uniqueness in quasar absorption models and implications for measurements of the fine structure constant}

\author[C.C. Lee et al]{
Chung-Chi Lee$^1$\thanks{E-mail: lee.chungchi16@gmail.com},
John K. Webb$^1$\thanks{E-mail: jkw.phys@gmail.com},
Dinko Milakovi{\'c}$^2$,
and Robert F. Carswell$^3$.\\\\
$^1$Clare Hall, University of Cambridge, Herschel Rd, Cambridge CB3 9AL.\\
$^2${Institute for Fundamental Physics of the Universe, Via Beirut, 2, 34151 Grignano TS, Italy.}\\
$^3${Institute of Astronomy, University of Cambridge, Madingley Road, Cambridge CB3 0HA, U.K.}
}

\date{Received in original form: 23 February 2021.\\ Accepted:}
\pubyear{2021}
\begin{document}
\label{firstpage}
\pagerange{\pageref{firstpage}--\pageref{lastpage}}
\maketitle

\begin{abstract}
High resolution spectra of quasar absorption systems provide the best constraints on temporal or spatial changes of fundamental constants in the early universe. An important systematic that has never before been quantified concerns model non-uniqueness. The absorption structure is generally complicated, comprising many blended lines. This characteristic means any given system can be fitted equally well by many slightly different models, each having a different value of $\alpha$, the fine structure constant. We use AI Monte Carlo modelling to quantify non-uniqueness. Extensive supercomputer calculations are reported, revealing new systematic effects that guide future analyses: (i) Whilst higher signal to noise and improved spectral resolution produces a smaller statistical uncertainty for $\alpha$, model non-uniqueness adds a significant additional uncertainty. (ii) Non-uniqueness depends on the line broadening mechanism used. We show that modelling the spectral data using turbulent line broadening results in far greater non-uniqueness, hence this should no longer be done. Instead, for varying $\alpha$ studies, it is important to use the more physically appropriate compound broadening. (iii) We have studied two absorption systems in detail. Generalising thus requires caution. Nevertheless, if non-uniqueness is present in all or most quasar absorption systems, it seems unavoidable that attempts to determine the existence (or non-existence) of spacetime variations of fundamental constants is best approached using a statistical sample.

\end{abstract}


\section{Introduction}

Generalised theories of varying fundamental constants \citep{Barrow2012} motivate high-precision searches for new physics using new facilities like the Echelle SPectrograph for Rocky Exoplanets and Stable Spectroscopic Observations (ESPRESSO) on the European Southern Observatory's Very Large Telescope (VLT) \citep{espresso2021}. Varying constants constitute one of the main scientific drivers for the forthcoming Extremely Large Telescope (ELT) e.g. \citep{Marconi2016, ELT2018}. Instrumentation improvements, better wavelength calibration methods based around laser frequency combs \citep{Milakovic2020, Probst2020}, and ever-increasing data quality, necessitate a new look at old analysis methods; approximations made when data quality was lower may be inadequate today.

One such approximation arises in the context of modelling quasar absorption systems. Minimising the number of free model parameters is appealing, to reduce degeneracy between parameters and because it generally yields a smaller uncertainty estimate on the parameter or parameters of interest. However, in the interests of objectivity, best-fit models should be free from human decision-making \citep{Lee2020AI-VPFIT} and the number of model parameters should be information criterion-based \citep{Webb2021}.

Nevertheless, even with fully objective and reproducible methodologies, model ambiguity is unavoidable; the $\chi^2$--parameter space may contain multiple local minima and more than one set of model parameters may provide a statistically acceptable fit to the data. This property generates an additional uncertainty on interesting parameters, over and above the usual ``statistical'' covariance matrix uncertainty derived at the minimum $\chi^2$. Additional uncertainties of this sort increase data scatter, sometimes allowed for by ``$\sigma_{\textrm{rand}}$'' in statistical surveys of the fine structure constant, $\alpha$, at high redshift, e.g. \citep{Webb2011}.

For varying $\alpha$ searches using quasar absorption systems, it is necessary to assume a broadening mechanism for each redshift component in an absorption complex. In this paper, we use an extensive suite of supercomputer simulations to study in detail how the assumed broadening mechanism and the choice of information criterion influences the parameter error budget and model ambiguity.

\section{Artificial Intelligence Monte Carlo} \label{sec:AIMC}

\subsection{Why AIMC is needed}

In \citep{Lee2020AI-VPFIT} we introduced an artificial intelligence approach to modelling absorption line systems ({\sc ai-vpfit}), a development built on the first attempt at such a code, described in \citep{Bainbridge2017, gvpfit17}. {\sc ai-vpfit} merges a genetic algorithm, a Monte Carlo process, and a well-established code, {\sc vpfit} \citep{VPFIT,web:VPFIT} to form a fully automated and unbiased modelling process. The statistical error returned by {\sc vpfit} is derived from the Hessian diagonal, based on a parabolic model for the $\chi^2$-parameter space. It has been shown that the {\sc vpfit} error estimates are reasonable e.g. \citep{King2009}. Nevertheless, the parabolic model cannot, of course, map out $\chi^2$ to reveal potential multiple minima. When multiple minima are present, the scatter we see amongst a distribution of {\sc ai-vpfit} models should be larger than the statistical error. 

The Monte Carlo aspect of {\sc ai-vpfit} (trial absorption line positions are placed randomly and repeatedly in an iterative procedure) means that each independent model is constructed differently; parent models for each generation are not the same between independent {\sc ai-vpfit} models. This slight procedural variation emuates the slightly different approaches different human modellers would take, presented with the same problem. The main difference is, of course, that a human may require weeks to carry out the same task that {\sc ai-vpfit} carries out in hours. This procedural variation in turn means that all (or most) multiple minima present in $\chi^2$-parameter space should be revealed in the calculations reported here, given enough independent {\sc ai-vpfit} runs, as would happen in a Markov-Chain Monte Carlo approach \citep{King2009}. 

If the line broadening model is wrong (as is often likely to be the case for pure turbulent or thermal broadening models), it would be expected that a broader range of best-fit models can be found. To give a simple example, if a turbulent model is imposed on an observed line that is in reality thermally broadened, the model line width for that line could potentially be (for example) too small (depending on atomic mass and other things). In this case, additional spurious velocity components may be added which can then velocity-shift components of species sensitive to $\daa$ and hence bias the measurement. The likely consequence is that additional components will be needed for that particular line in order to obtain a good fit. Therefore, the model non-uniqueness problem should be greater if we use a pure turbulent or a pure thermal model, compared to compound broadening, which accommodates both thermal and turbulent mechanisms.

Whilst the $\chi^2$ statistic is useful in helping to select a ``best-fit'' theoretical model to any absorption complex, that statistic provides little guidance as to whether too many or too few model parameters have been used. For that reason, it is useful to use an {\it Information criterion}, which takes the general form
\begin{equation}
\rm{IC} = \chi^2(N_p,N_d) + \mathcal{P}(N_p,N_d)
\label{eq:IC}
\end{equation}
where $\rm N_p$ is the number of model parameters and $\rm N_d$ is the total number of data points. The first term in Eq.\eqref{eq:IC} has the usual definition. The second term is a penalty term that increases with increasing number of model parameters. The second term thus regulates the IC such that it minimises for a best-fit model (whereas $\chi^2$ cannot minimise as a function of the number of free parameters). The best-fit model is thus taken to be that for which the IC minimises. Different ICs take different forms for the penalty term $\mathcal{P}(\rm N_p,N_d)$, so different ICs may not necessarily agree on the best-fit model. This issue was recently studied in detail in \cite{Webb2021}. That paper introduces a new information criterion (SpIC) aimed specifically at spectroscopic modelling, to address weaknesses identified (in this context) for the widely used information criteria, the corrected Akaike Information Crition \citep{Akaike1974, Hurvich1989}, 
\begin{equation}\label{eq:AICc}
\mathrm{AICc} = \chi^2 + \frac{2 N_p N_d }{N_d - N_p -1}\,\,,
\end{equation}
and the Bayesian Information Criterion, \citep{Bozdogan1987}, (BIC). The finding of that work was that SpIC ``bridged the gap'' between the overfitting and underfitting tendencies of AICc and BIC respectively \citep{Webb2021}. With the availability of a system such as {\sc ai-vpfit}, the considerations above motivate a more detailed exploration of $\chi^2$-parameter space using the various broadening models as well as using different information criteria. SpIC is defined as
\begin{equation}
\label{eq:SpIC}
\mathrm{SpIC} = \chi^2 + \sum_{a=1}^Q \left[ \frac{ k_a R_a }{R_a - k_a -1} + \frac{k_a\ln(R_a)}{2} \right]\,\,,
\end{equation}
where $Q$ is the total number of velocity components in the absorption model, $k_a$ is the number of free parameters associated with each velocity component such that $N_p = \sum k_a$, $R_a$ is the effective total absorption strength, defined as
\begin{equation} \label{eq:Ra}
R_a = \sum_{j=1}^M \sum_{i=1}^{N_j} \left( \frac{ 1-I_{i,j}^a}{ \sigma_{i,j}}\right) \,,
\end{equation}
where $M$ is the number of spectral segments, $N_j$ is the number of pixels in $j^{th}$ spectral segment, $I_{i,j}$ is the normalised spectral intensity from the model fit for $i^{th}$ pixel in $j^{th}$ spectral segment, and $\sigma_{i,j}$ is the corresponding uncertainty from the observed spectrum. Equations \eqref{eq:AICc} and \eqref{eq:SpIC} are used in the present study to compare the degree of model non-uniqueness obtained in the {\sc ai-vpfit} model fitting procedure when using these two statistics.

\subsection{Astronomical data} \label{subsec:2systems}

The analyses described in this study make use of two quasar spectra. The first is is a segment within the complex system at $z_{abs}=1.147$ towards the bright quasar HE0515$-$4414 (emission redshift $z_{em}=1.713$, V magnitude 15.16). The data were obtained using the HARPS spectrograph on the ESO 3.6m telescope and were wavelength calibrated using a Laser Frequency Comb. The spectral resolution (FWHM) obtained was $R = \lambda/\Delta\lambda \approx 115,000$, higher than the majority of published quasar echelle spectra. The average signal to noise per pixel is around 50.

Since the present work requires considerable computing power (models contain a large number of free parameters), we select a subset of the data: the redshift range used is $1.14658 < z_{\textrm{abs}} < 1.14778$, corresponding to 167 km/s, about 1/4 of the whole absorption complex. Restricting the range in this way serves two purposes: the range is representative of typical quasar absorption systems and it also helps to reduce the required computing time to an acceptable level.

A detailed description of the HE0515$-$4414 spectrum used is given in \citep{Milakovic2021} and we refer the reader to that paper for further details. A detailed study of this quasar, using lower resolution data obtained using the Ultra Violet Echelle Spectrograph (UVES) on the VLT, is reported in \cite{Kotus2017}.

The intrinsic characteristics of the $z_{abs}=1.147$ absorption system towards HE0515$-$4414 (or section thereof) analysed in the previous section is typical of many others. The resolution of the HARPS spectrum ($R \approx 115,000$ FWHM) is typical of new data from instruments like ESPRESSO on the VLT and of data to come from facilities like the High Resolution Echelle Spectograph (HIRES) on the ELT. However, it is atypical of much of the quasar data available at the time of writing this paper; most of the currently available quasar spectra from UVES on the VLT or HIRES on the Keck telescope have a lower spectral resolution, typically $R \approx 50,000$ FWHM. This raises the question as to how general the results from HE0515$-$4414 are. Therefore we repeat the analysis of the previous section using a second absorption system, the $z_{abs}=2.141$ system towards the well-known quasar Q0528$-$2505 (or J053007-250329). This quasar has an emission redshift $z_{em}=2.765$ and a V magnitude of 17.34. The absorption complex spans approximately 260 km/s and the signal to noise per pixel is approximately 65 at 5000{\AA}, reducing to about 23 at 8800{\AA}. The spectral resolution is $R \approx 50,000$ FWHM. The spectrum was obtained using UVES on VLT and further observational details are given in \cite{Murphy2019}. The data quality and characteristics of this absorption system are representative of much of the existing quasar data.

\subsection{Line broadening}

For pure thermal broadening, the width of an absorption line depends only on the cloud temperature $T$ and atomic mass, $m$. However, for sufficiently low cloud temperatures, line broadening can be dominated by bulk motions in the cloud. More generally, we may find that both processes apply, such that line widths depend on three parameters, $T$, $m$, and an additional turbulence parameter. Here we adopt the usual assumption that the observed line profile is Voigt so that thermal and turbulent contributions can be combined in quadrature and thus study the following three cases:\\
\noindent 1. Fully turbulent broadening; all atomic species at the same redshift share the same $b$-parameter, $b_{\textrm{turb}}$,\\
\noindent 2. Pure thermal broadening,  $b_{\textrm{th}}=\sqrt{\frac{2kT}{m}}=12.85\sqrt{\frac{T}{10^4 m}}$ km s$^{-1}$,\\
\noindent 3. Compound broadening, $b_{\textrm{obs}}^2 = b^2_{\textrm{turb}} + b_{\textrm{th}}^2$.\\
Therefore, when fitting an absorption component using compound broadening, one addition parameter is required compared to the turbulent and thermal case.

\subsection{Algorithm stopping criteria}

Non-linear least squares algorithms such as {\sc vpfit} (which is incorporated into {\sc ai-vpfit}) require carefully-set stopping criteria to terminate the minimisation process. The usual approach is to set $\Delta\chi^2 / \chi^2 = (\chi^2_n - \chi^2_{n-1})/\chi^2_{n-1}$ (where $n$ indicates the iteration number) to be smaller than some suitable threshold, such that any parameter changes between the final two iterations are well below their corresponding 1$\sigma$ uncertainties. For the calculations described in this paper, we set $\Delta\chi^2 / \chi^2 = 5\times 10^{-4}$. This ensures a very small increment between successive values of $\daa$, certainly far below its statistical uncertainty. For HE0515$-$4414, this stopping criteria corresponds to $\Delta\chi^2 \lesssim 0.75$, which maps (empirically) to a change $|\daa| \lesssim 2 \times 10^{-7}$, approximately an order of magnitude below the estimated statistical uncertainty on $\daa$. As a precaution, we set the requirement that the stopping criterion must be met on three consecutive iterations before terminating the fit. We can thus be confident that each of the {\sc ai-vpfit} models correspond closely to a real local minimum in $\chi^2$ space.

\subsection{High Performance Computing}

The high performance supercomputing facility used for the calculations in this paper is OzSTAR at Swinburne University in Australia\footnote{\url{https://supercomputing.swin.edu.au/ozstar/}}. The computing time required for automated modelling using a procedure such as {\sc ai-vpfit} is significant. For this reason, several parts of the code have been parallelised to run simultaneously across multiple processors. Even with parallelisation, approximately 80,000 computing hours were required on OzSTAR to obtain the HE0515$-$4414 results presented in this paper. We generate a total of about 600 models, modelling a total of 8 spectral transitions from 3 atomic species. The quasar spectrum is fitted independently 100 times, using 3 line broadening models and 2 information criteria, AICc and SpIC \citep{Webb2021}, to select the fittest model at each generation. Both ICs have been incorporated into the {\sc ai-vpfit} code (the SpIC modification to {\sc ai-vpfit} is not described in \citep{Lee2020AI-VPFIT} since that enhancement has since been added). On average, each model thus requires 135 hours processing time. Parallelisation (we typically used 5 CUPs per calculation) then means the average model calculation time is 27 hours for HE0515$-$4414.

The details are slightly different for the second absorption system, $z_{abs}=2.141$ towards Q0528-250. In this case we have 12 useful atomic transitions from 6 species. As before, we carry out HPC calculations using OzSTAR to model the absorption system multiple times. Calculations were made using both AICc and SpIC information criteria, for turbulent, compound, and thermal broadening models. For each configuration we generated 50 models i.e. a total of 300 models were computed. Using the OzSTAR facility these calculations took a grand total of approximately 120,000 computing hours. The longer computing time for this second system is due to the larger number of transitions fitted and the larger velocity spread of the absorption system.

\begin{figure*}
\centering \vspace{-1cm}
\includegraphics[width=1.0\linewidth]{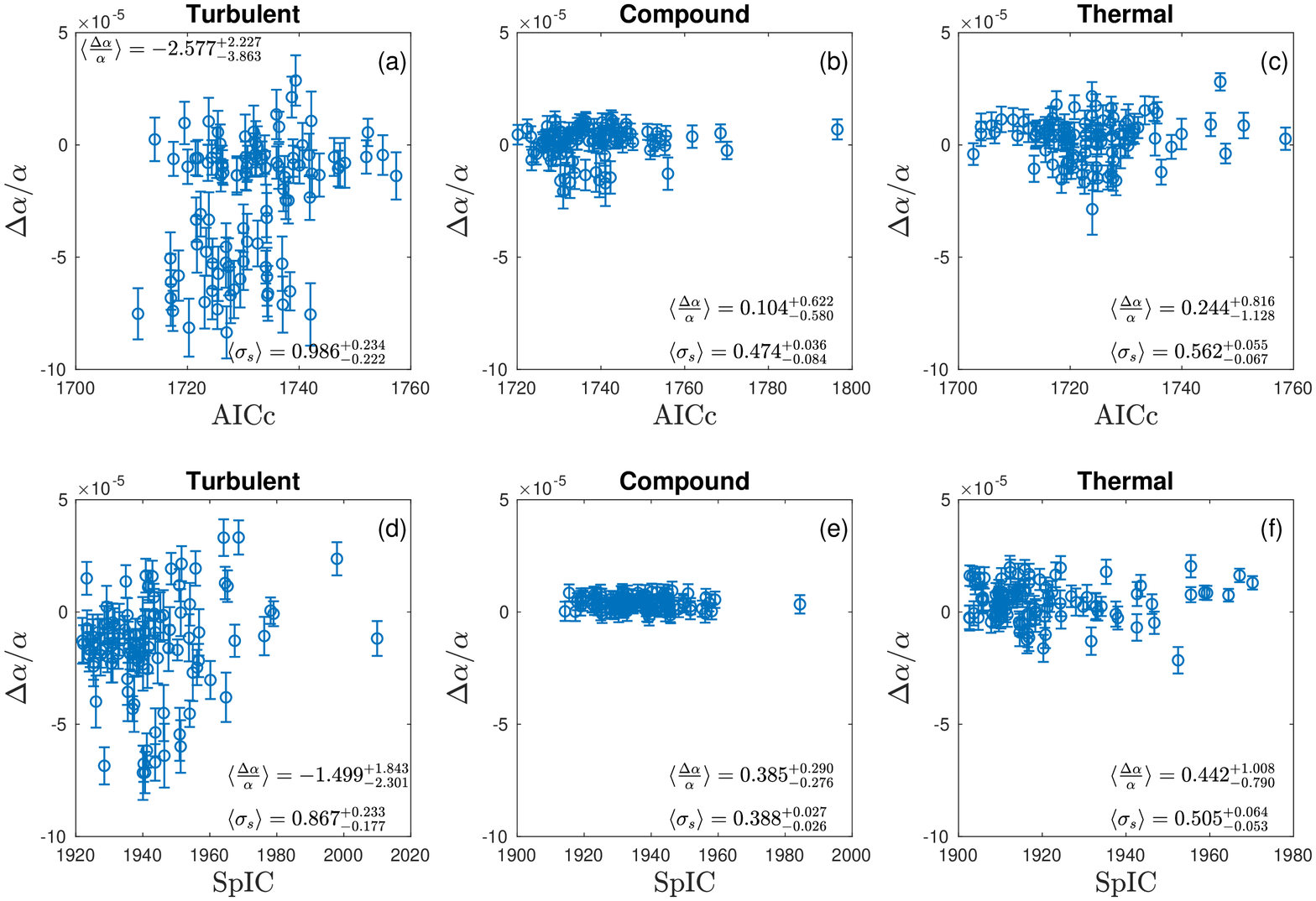} \\ 
\includegraphics[width=1.0\linewidth]{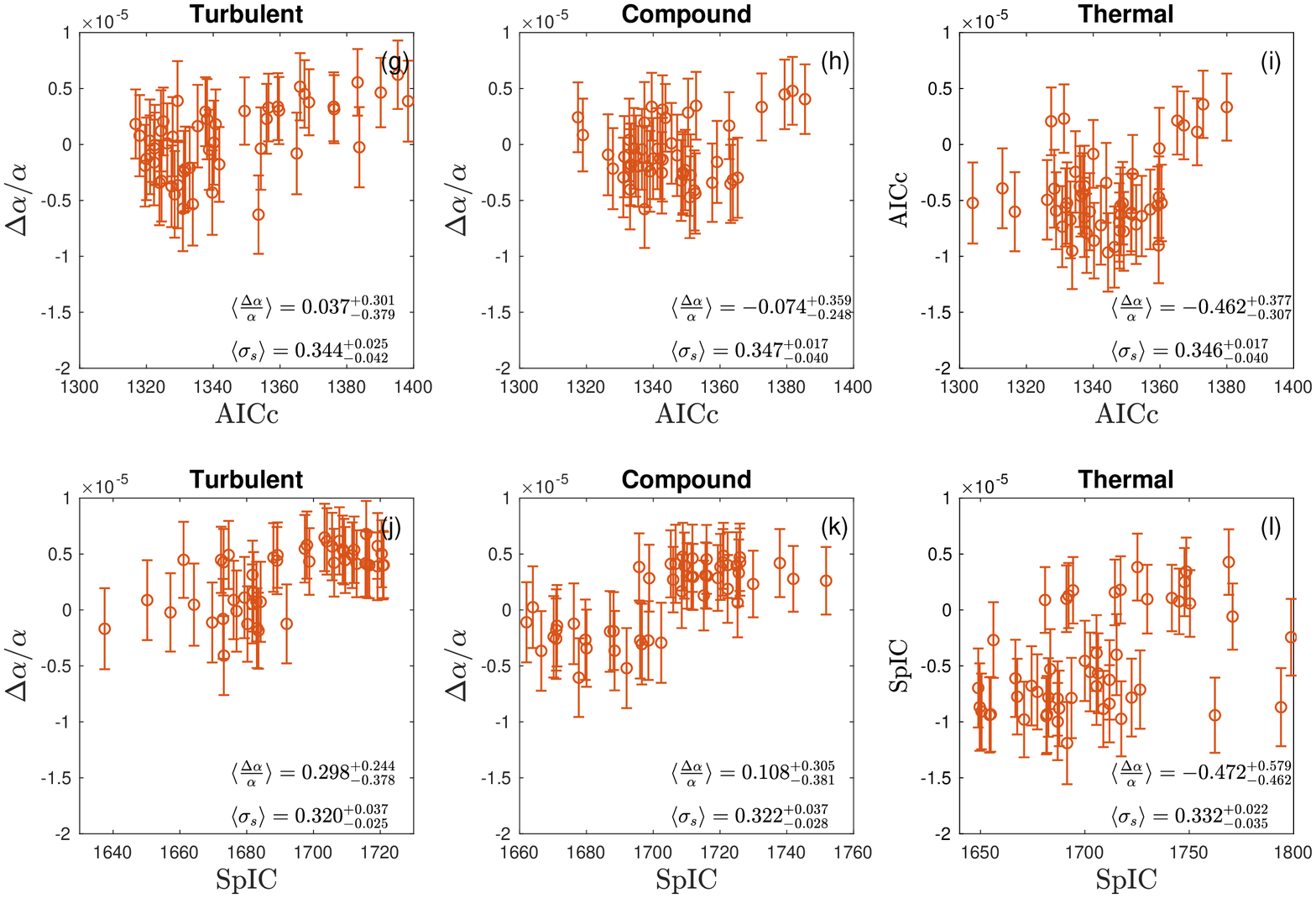} \\
\caption{$\daa$ (in units of $10^{-5}$) vs both information criteria, AICc and SpIC. The two top rows give results for HE0515$-$4414 (blue) and the lower two rows are for Q0528$-$2505 (red). Row 1 is for AICc, row 2 for SpIC, row 3 for AICc, and row 4 for SpIC. Approximately 100 models were computed for each of the HE0515$-$4414 panels whilst 50 were computed for each of the Q0528$-$2505 panels. Each hollow circle corresponds to one {\sc ai-vpfit} model The error bars plotted are from the {\sc vpfit} Hessian diagonal. Within each panel we show the simple mean $\daa$ and its 68\% range super- and sub-scripts for the different line broadening models. The 68\% range does not include the statistical uncertainty $\sigma_{s}$ i.e. the uncertainty returned by the {\sc vpfit} Hessian diagonal, so the range illustrated can be considered to represent an {\it additional} systematic error associated with model non-uniqueness, over and above the statistical uncertainty. Also within each panel we show the mean statistical uncertainty $\langle \sigma_{s} \rangle$.}
\label{fig:6x2ICs}
\end{figure*}

\begin{figure*}
\centering \vspace{-1cm}
\includegraphics[width=1.0\linewidth]{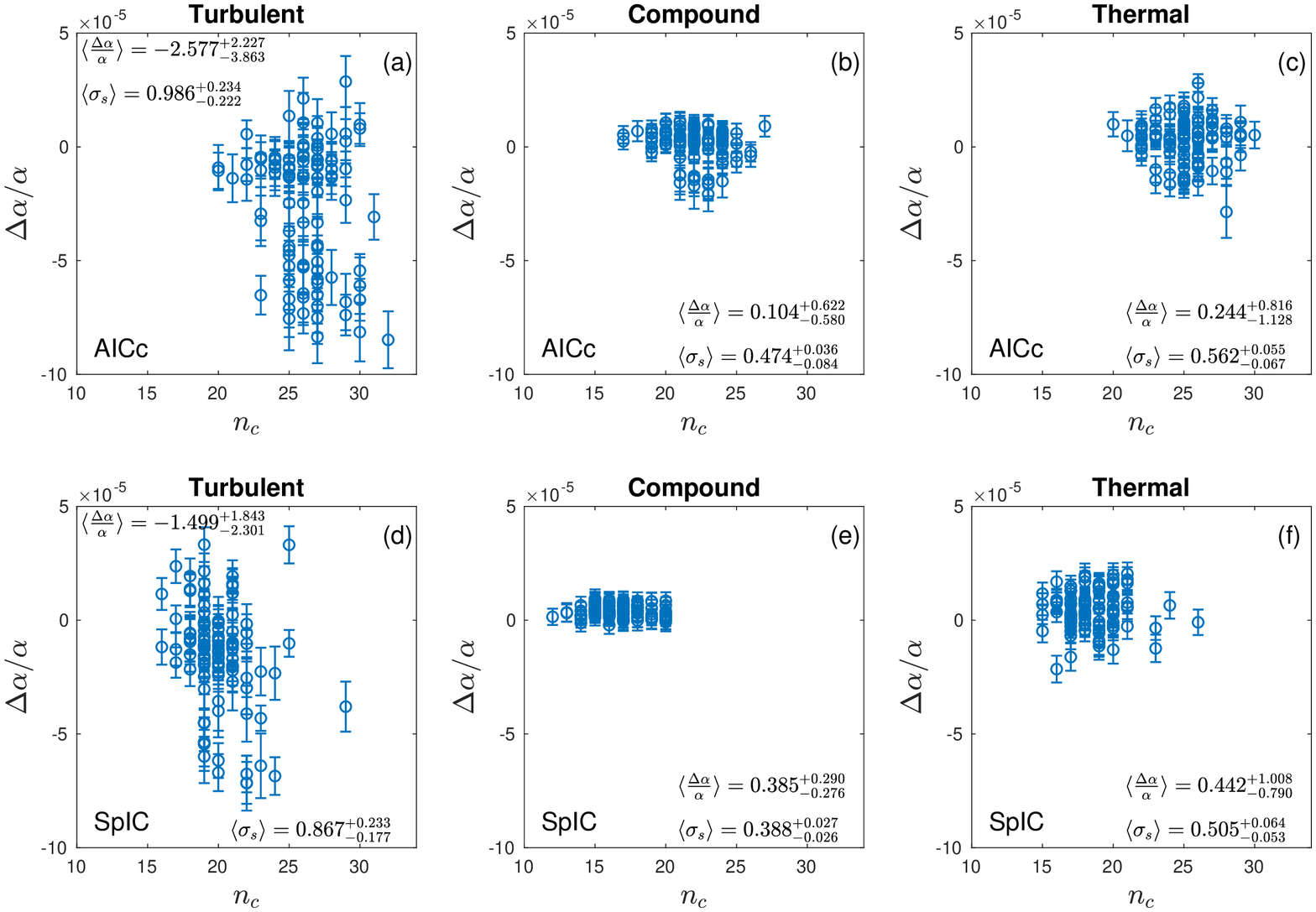} \\ 
\includegraphics[width=1.0\linewidth]{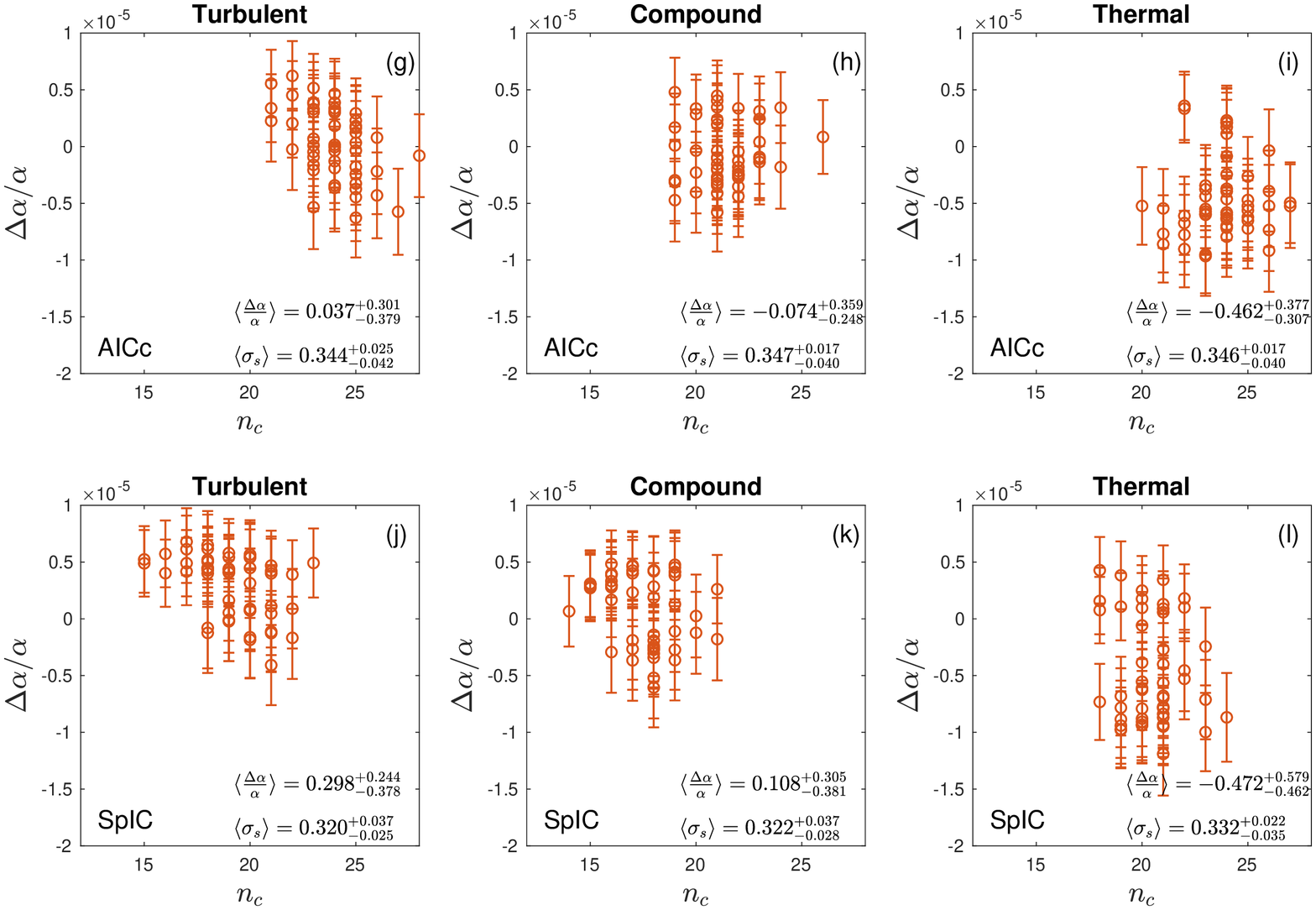} \\ 
\caption{$\daa$ (in units of $10^{-5}$) vs. $n_c$, the number of model components, including both metal and interloper lines. All panels correspond to Figure \ref{fig:6x2ICs}. The plots illustrate that SpIC requires fewer model components compared to AICc.}
\label{fig:6x2nc}
\end{figure*}

\begin{figure*}
\centering \vspace{-1cm}
\includegraphics[width=1.0\linewidth]{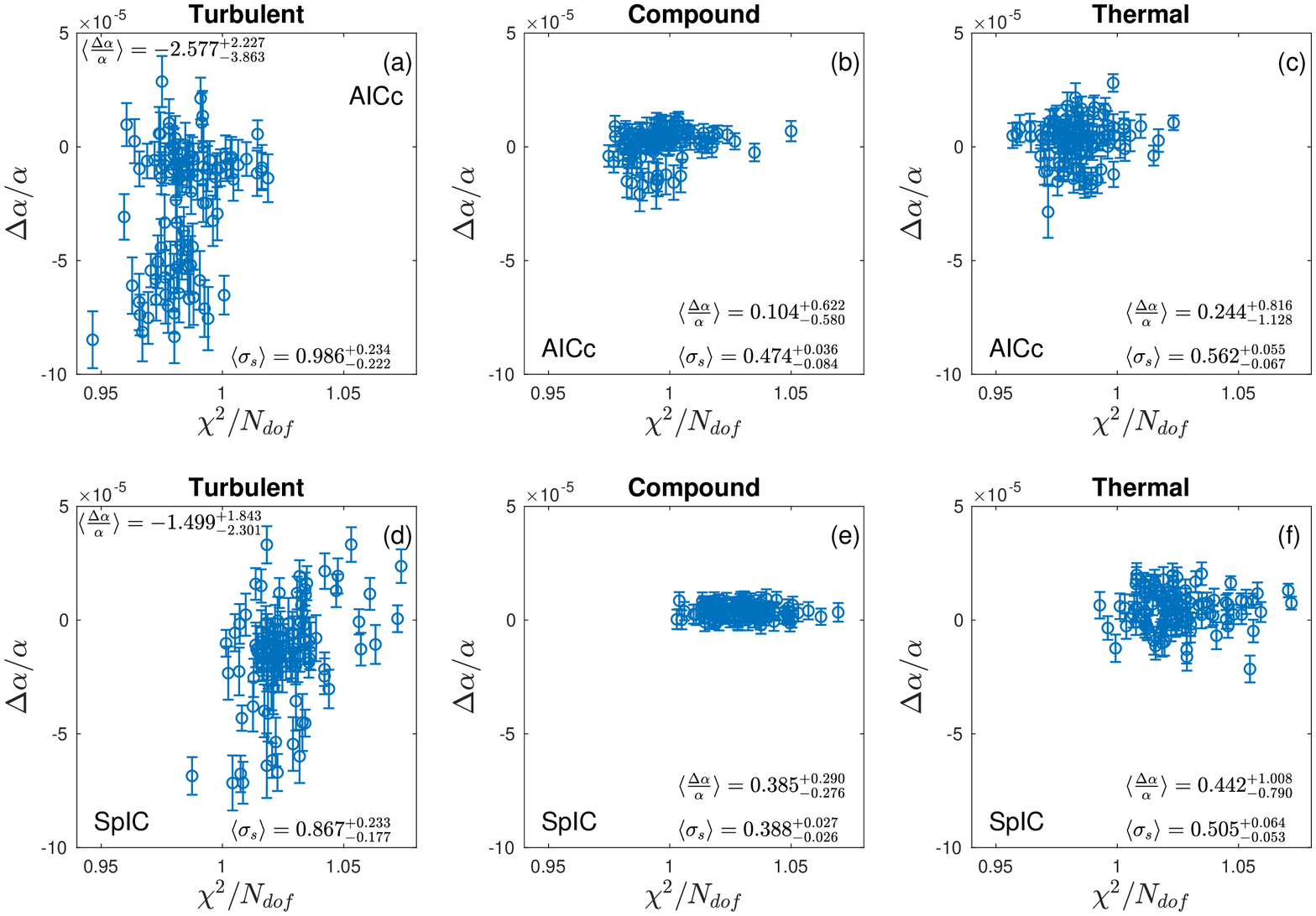} \\ 
\includegraphics[width=1.0\linewidth]{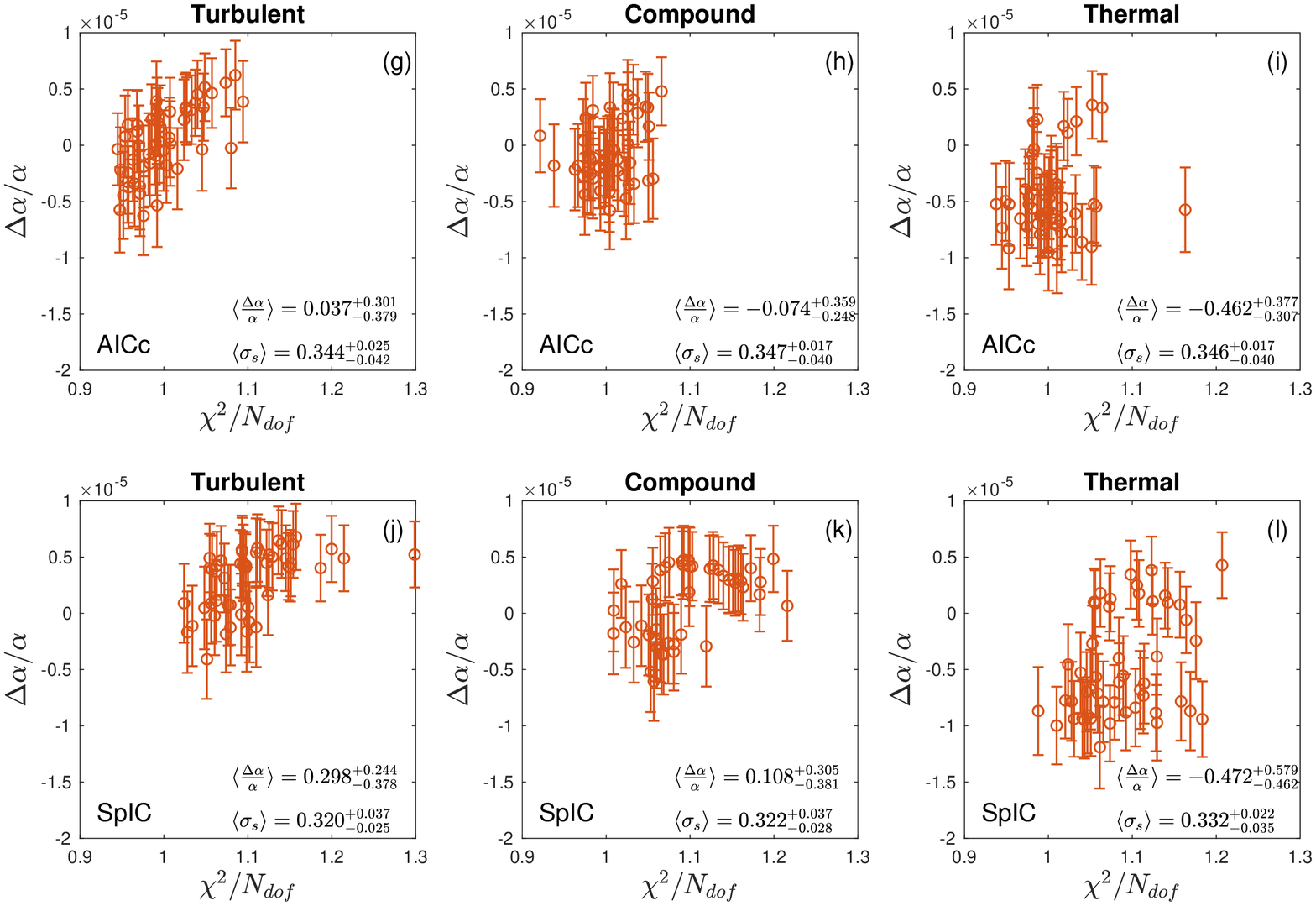} \\
\caption{$\daa$ (in units of $10^{-5}$) vs. $\chi^2 / N_{dof}$, where $N_{dof} = N_d - N_p$, the difference between the total number of data points and the number of model parameters. All panels correspond to Figure \ref{fig:6x2ICs}. This figure shows that AICc generally produces a slightly smaller value of $\chi^2 / N_{dof}$, as expected, and that both information criteria produce statistically acceptable models (i.e. $\chi^2 / N_{dof} \approx 1$ in all cases). The results from HE0515$-$4414 (upper two rows, blue points) indicate that non-uniqueness is a more significant problem for turbulent models and also for AICc compared to SpIC. Those distinctions are masked in the lower resolution spectrum of Q0528$-$2505 (lower two rows, red points).}
\label{fig:6x2chisq}
\end{figure*}

\section{Analysis of HE0515$-$4414} \label{sec:AnalysisHE0515}

The distribution of models we get is unbiased in that each model is constructed using a genetic procedure in which first guesses are generated using a Monte Carlo method \citep{Lee2020AI-VPFIT}. As we will shortly see, a new source of uncertainty is revealed, over and above the simple ``statistical'' uncertainty derived from the Hessian diagonals at the best-fit solution. 

Figures \ref{fig:6x2ICs}, \ref{fig:6x2nc}, and \ref{fig:6x2chisq} illustrate the results obtained. The HE0515$-$4414 absorption system has been fitted with {\sc ai-vpfit} about 100 times each, using turbulent, thermal, and compound line broadening models and using the two information criteria, SpIC \citep{Webb2021} and AICc (i.e. making a total of $\sim$600 independent models). Each panel in these Figures shows the distribution of best-fit $\daa$ values. Each point corresponds to a different {\sc ai-vpfit} model. Since the model construction process is Monte Carlo based \citep{Lee2020AI-VPFIT}, the observed distribution of points will reveal multiple minima, if present.

In Figure \ref{fig:6x2ICs}, panels (a) and (d) reveal something quite remarkable. These panels correspond to turbulent line broadening, which is the model frequently assumed when modelling quasar absorption systems. Whilst a turbulent model is recognised as an approximation, it has previously been assumed that the assumption does not introduce any bias in estimating $\daa$ and that it does not add any additional uncertainty to the measurement. The turbulent broadening panels illustrate that both assumptions are wrong. Consider again panel (a): the AICc models fall into two well-separated clumps, one at around $\daa \approx -1$, the other around $\daa \approx -6$ (in units of $10^{-5}$). The upper clump is populated by around 60\% of the best-fit models whilst the lower clump is populated by around 40\%. If using AICc, we do not know, {\it a priori}, which of these is correct. A single model, produced interactively by a human, is expected to suffer from these difficulties. The overall spread in $\daa$ is actually {\it substantially larger than the statistical uncertainty returned by the Hessian diagonal}. These results are both surprising and somewhat shocking, given many measurements in the literature are based on the process just outlined. In fact, Figure \ref{fig:6x2ICs} shows that the worst possible assumption, at least for HE0515$-$4414, is turbulent broadening. The ``non-uniqueness'' problem is far less pronounced for thermal, although compound broadening is clearly the preferable model.

More generally, Figure \ref{fig:6x2ICs} shows that if the best-fit model $\chi^2$-parameter space contained one single global minimum and was otherwise structureless, all {\sc ai-vpfit} models would be identical. That they are not is a demonstration of real multiple local minima. Each panel shows the simple mean and range for $\daa$ (the range is determined empirically by the $\pm 34$\% range over the sample of $\sim$100 models in each case). The mean statistical uncertainty using AICc is $\langle \sigma_s \rangle = 0.986 \times 10^{-5}$ for the turbulent case, compared with $0.474 \times 10^{-5}$ for the compound broadening model. The corresponding numbers for SpIC are smaller but show the same trend. These numbers already demonstrate convincingly that SpIC is preferable to AICc and that the appropriate broadening model is compound, not turbulent and not thermal.

Prior to a code such as {\sc ai-vpfit} \citep{Lee2020AI-VPFIT}, the calculations described here would have been impractical. However we now learn something interesting: for both SpIC and AICc, the scatter in the turbulent samples is large compared to the statistical error. This is of considerable concern because the systematic error associated with model non-uniqueness appears (in the HPC calculations) to be considerably larger than the statistical error. This may be interpreted in two possible ways: either (a) turbulent models represent the data badly and hence generate multiple $\chi^2$--parameter space minima, or (b) the Monte Carlo nature of the {\sc ai-vpfit} modelling process (i.e. placing trial lines {\it randomly} within the spectral segment being fitted) may not emulate a human process and may itself generate multiple $\chi^2$--parameter space minima that are particular to {\sc ai-vpfit}. If the explanation is simply (a), we may expect to see the same effect in published $\daa$ samples that were fitting using a turbulent model. Although both {\sc ai-vpfit} and a human interactive modeller will tend to target strong but unsaturated features earlier and refine the model with weaker features later, there is no ``correct'' ordering in which an absorption complex model should be constructed. Any clumping in $\chi^2$ space is not an artificial aspect of the {\sc ai-vpfit} Monte Carlo process, which itself removes any possible subjectivity.

In Figure \ref{fig:6x2nc} we show how the different broadening models and the two different information criteria impact on the required number of model parameters. A clear trend is seen; SpIC  requires fewer model parameters. Given the ways in which AICc and SpIC are defined, this is expected. We refer to \cite{Webb2021} for a more detailed discussion on this point.

Figure \ref{fig:6x2nc} reveals something else rather interesting; AICc and SpIC generate quite different solutions for the HE0515$-$4414 absorption system, irrespective of the broadening method. The AICc compound broadening model results in a mean number of parameters of roughly 120 whereas the SpIC compound broadening model produces about 100. This means the AICc models, on average, require around 7 more absorption lines across the complex. Closer inspection of the results shows that these divide approximately equally into heavy element components and interlopers.

Figure \ref{fig:6x2chisq} reinforces the above as we can see that on average, the overall best-fit $\chi^2$ values are slightly smaller for AICc compared to SpIC, as expected. Both information criteria provide ``statistically acceptable'' model fits i.e. the normalised values of $\chi^2$ are around unity. Trying to distinguish between these possibilities using $\chi^2$ is not reliable because spectral processing procedures from raw data to one dimensional spectrum create weak small-scale pixel to pixel correlations. This means that the spectral error array is only an approximation and small departures of the normalised $\chi^2$ from unity are not easy to interpret.

\begin{figure}
\centering
\includegraphics[width=1.0\linewidth]{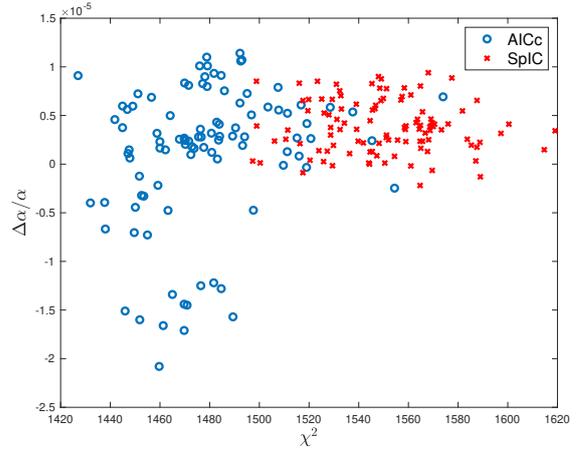}
\caption{$\daa$ for HE0515$-$4414 from AICc {\sc ai-vpfit} models (blue circles) and SpIC (red crosses) as a function of $\chi^2$. The line broadening is compound. This figure corresponds to panels (b) and (e) in Figures \ref{fig:6x2ICs}, \ref{fig:6x2nc}, and \ref{fig:6x2chisq}. The overfitting properties of AICc generate the $\chi^2$ sub-structure and model non-uniqueness dominates the overall $\daa$ uncertainty.}
\label{fig:clump}
\end{figure}

\begin{figure*}
\centering
\includegraphics[width=1.0\linewidth]{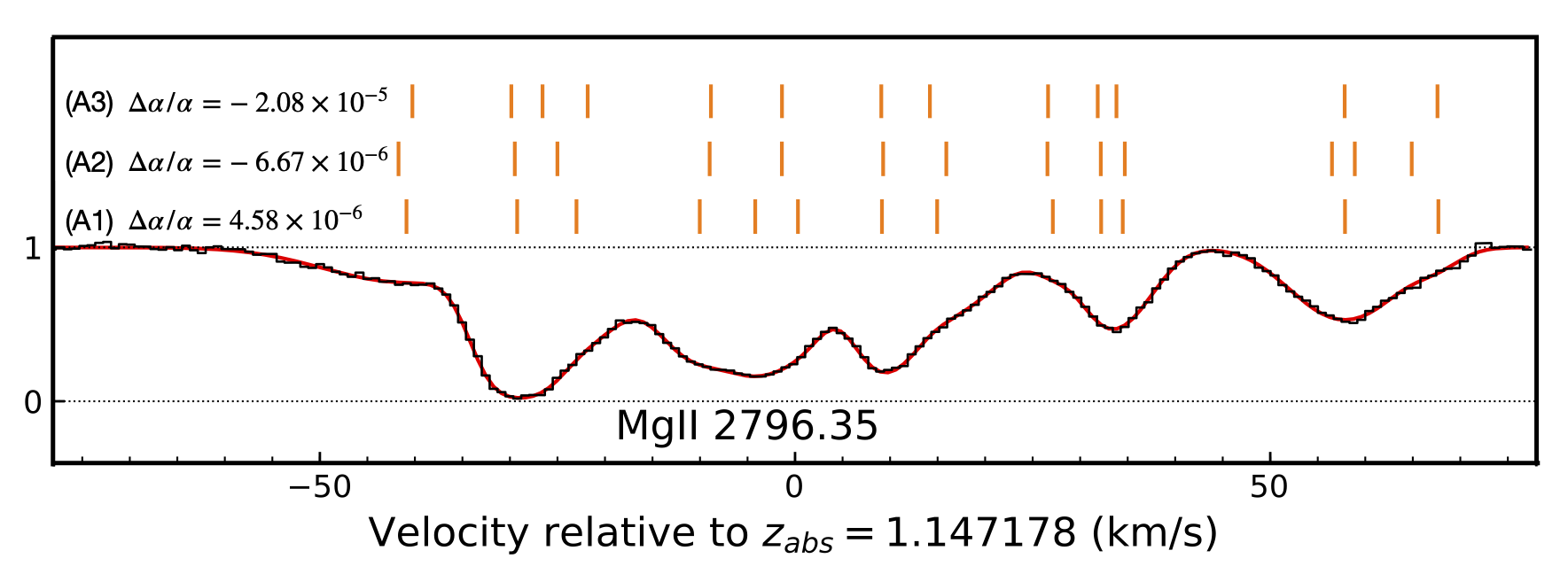}
\caption{Three example {\sc ai-vpfit} models (using AICc) for a segment of the absorption complex towards HE0515$-$4414. These are compound broadening fits (i.e. the lines are broadened by both thermal and turbulent Doppler motions). Each row of tick marks illustrate the component positions for each of the three models and the corresponding best-fit $\daa$ values (more detailed figures for each model are given in the Appendix). Model A1 has $\daa = (-20.8 \pm 7.52) \times 10^{-6}$, model A2 $\daa = (-6.67 \pm 4.60) \times 10^{-6}$, and model A3 has $\daa = (4.58 \pm 4.38) \times 10^{-6}$.}
\label{fig:3models}
\end{figure*}

The blue hollow circles in Fig.~\ref{fig:clump} illustrate $\daa$ vs the normalised $\chi^2$ for 98 model fits using AICc. These models are for compound line broadening. The red crosses show 100 fits using SpIC. The AICc distribution appears to fall into three (or possibly four) clumps: the most populated lies in the approximate range $0 < \daa < 1 \times 10^{-5}$. Two others lie in the approximate range $-0.75 \times 10^{-5} < \daa < 0$ and $-2.1 \times 10^{-5} < \daa < -1.1 \times 10^{-5}$. As can be seen, the spread in $\daa$ is $\sim 3 \times 10^{-5}$, approximately 6 times larger than the statistical ({\sc vpfit}) uncertainty on each individual point (Fig.~\ref{fig:6x2chisq}). The blue circles thus emphasise that for AICc models specifically, non-uniqueness is a highly significant issue, even when the correct line broadening model is used. To see this further, Figure \ref{fig:3models} shows three example models, drawn {\it ad hoc} from each of the three clumps. All three models produce very similar $\chi^2$ values and visually indistinguishable models and normalised residuals (details are shown in Table \ref{tab:3models} and all transitions modelled are shown in Figures \ref{fig:spec_HE0515a}, \ref{fig:spec_HE0515b}, and \ref{fig:spec_HE0515c}). If modelling is carried our interactively by a human, it is unclear where in the $\daa$-$\chi^2_{\nu}$ plane the fit might end up. Of course it is possible that an interactive modeller might be tempted (either consciously or otherwise) to manually adjust parameters that give $\daa$ closest to zero.

\begin{table}
\begin{tabular}{l l l l } 
\hline 
Model: & \textbf{A1} & \textbf{A2} & \textbf{A3}  \\
 \hline
Metals & 13 & 13 & 13 \\
$\daa (10^{-6})$ & $-20.8 \pm 7.52$ & $-6.67 \pm 4.60$ & $ 4.58 \pm 4.38$ \\
$\chi^2_{\nu}$ & 0.9876 & 0.9769 & 0.9775 \\
AICc & 1730.997 & 1723.466 & 1720.158 \\
\hline
\end{tabular}
\caption{Statistical details for AICc models A1, A2, A3, as illustrated in Figs.~\ref{fig:spec_HE0515a}, \ref{fig:spec_HE0515b}, and \ref{fig:spec_HE0515c}. The first row (Metals) indicates the number of heavy element components in the fit.
\label{tab:3models}}
\end{table}

\section{Analysis of Q0528$-$2505} \label{sec:AnalysisQ0528}

The set of atomic transitions used in the model are shown in Figures \ref{fig:Q0528spec_a} and \ref{fig:Q0528spec_b}. The lower two rows, panels (g) through (l), in Figures \ref{fig:6x2ICs}, \ref{fig:6x2nc}, and \ref{fig:6x2chisq} illustrate the results for this absorption system. As with HE0515$-$4414, this second absorption system towards Q0528$-$2505 also exhibits non-uniqueness problems. Panel (h) in (for example) Figure \ref{fig:6x2ICs}, shows that the scatter amongst the 50 models calculated is $-0.248 < \daa < +0.359$ in units of $10^{-5}$ (this range encompasses 68\% of the $\daa$ values so is approximately equal to the standard deviation of the sample). The range is around twice the mean uncertainty derived from the best-fit Hessian diagonal. Whilst we do not see (for example) a clear bifurcation of the data points, this excess scatter over and above the ordinary statistical error implies that the $\chi^2$-$\daa$ space may not be smooth with one single minimum, but rather, is most likely complex in shape with multiple minima. In other words, non-uniqueness is present.

Panels (g) through (l) of Figure \ref{fig:6x2nc} show that using SpIC requires fewer velocity components compared to using AICc, irrespective of the broadening mechanism. As expected, the lower spectral resolution of the Q0528$-$2505 spectrum to some extent masks the advantage of compound broadening. Nevertheless, panels (g) and (j), and (i) and (l), all show multiple minima. These panels also show that the empirical scatter in $\daa$ is again significantly larger than the statistical (Hessian diagonal) uncertainty. Interestingly, for this absorption system, the same is true for compound broadening, panels (h) and (k), meaning that the advantage of SpIC over AICc seen for HE0515$-$4414 is also diminished by the lower spectral resolution. Alternatively, it is possible that the particular characteristics of the Q0528$-$2505 absorption system are such that SpIC and AICc are both equally effective - the data do not allow us to distinguish between these two possibilities, although the first seems more likely.

Interestingly, Figure \ref{fig:6x2chisq} panels (g) and (j) indicate clear trends. The explanation for this is clear from panels (g) and (j) in Figure \ref{fig:6x2ICs}; in those panels we see there are two strong local minima, one grouping at around $\daa = 0.5 \times 10^{-5}$, the other at around $\daa = 0$, such that the apparent trends in Figure \ref{fig:6x2chisq} are simply a consequence of non-uniqueness.

\section{Discussion} \label{sec:Discussion}

Modelling high resolution quasar absorption systems generally involves complex datasets and a large number of free parameters. It is well known that non-linear modelling procedures of this sort often generate complex $\chi^2$-parameters spaces i.e. multiple minima can exist, making interpretation of results less straightforward. The goal of this paper has been to explore in some detail the extent to which model non-uniqueness exists in previous analyses and the extent to which it may cause problems in future analyses.

To this end, we have analysed two quasar absorption systems using a new method that has not been applied before i.e. Artificial Intelligence Monte Carlo. The two absorption systems used are inherently representative but the two spectra have significantly different spectral resolutions; one spectrum (the $z_{abs}=1.147$ system towards the quasar HE0515$-$4414) has a resolution of $R \approx 115,000$ FWHM, typical of new and forthcoming data from instruments like ESPRESSO/VLT or HIRES/ELT. The other  (the $z_{abs}=2.141$ system towards the well-known quasar Q0528$-$2505) is more characteristic of the bulk of the existing high resolution quasar literature at a resolution of $R \approx 50,000$ FWHM. As would be expected, the higher resolution data provides greater insight to the issue of non-uniqueness, although both spectra reveal the same fundamental issues, as we now explain in more detail. We first discuss the findings from the HE0515$-$4414 spectrum and then compare those with the results obtained from the Q0528$-$2505 data.

\subsection{System 1: $z_{abs}=1.147$ towards HE0515$-$4414}

For the HE0515$-$4414 spectrum, our results show that fitting cloud models with turbulent line broadening generates or enhances model non-uniqueness, adding a substantial additional random uncertainty to any attempt to measure $\daa$. Many of the existing published $\daa$ measurements are based on (or involve) turbulent broadening so it is possible that this phenomenon contributes substantially to the excess scatter seen in previous samples of $\daa$ measurements\footnote{The excess scatters shown in Figures \ref{fig:6x2ICs}, \ref{fig:6x2nc}, and \ref{fig:6x2chisq}, over and above the statistical uncertainty, are comparable in magnitude and hence at least partially (and perhaps fully) account for excess scatter seen in the models of \citep{King2012}, parameterised using $\sigma_{rand}$.}.

Figures \ref{fig:6x2ICs}, \ref{fig:6x2nc}, and \ref{fig:6x2chisq}, panels (a) and (d), show how independent models clump into several groups, revealing multiple minima in $\chi^2$ space. As model complexity increases and additional components are introduced, relative line positions shift slightly generating different solutions for $\daa$. 

Panels (a) and (d) of (e.g.) Figure \ref{fig:6x2ICs} reveal the strikingly different behaviour of the AICc and SpIC statistics for HE0515$-$4414. The AICc measurements (panel (a)) bifurcate into two main groups, one around $\daa=0$, the other around $\daa= -6 \times 10^{-5}$. In contrast, the SpIC measurements strongly group around $\daa=-1 \times 10^{-5}$, with a weak but clear group around $\daa=-7 \times 10^{-5}$. The mean values of $\daa$ for each set of calculations are shown in the figure panels and (given these are not independent datasets) are inconsistent. The empirical $\daa$ scatter for AICc models is $(2.227+3.863) \times 10^{-5} \approx 6 \times 10^{-5}$, vastly higher than the mean statistical error (i.e. from the diagonal of the covariance matrix) $\langle \sigma_s \rangle = 0.986 \times 10^{-5}$. An empiricist, subjectively deriving one single model, may well preferentially select a model sitting in the $\daa=0$ grouping and conclude a job well-done, {\it but this is neither reasonable nor necessarily correct}.

The SpIC measurements for HE0515$-$4414 (panel (d) in Figure \ref{fig:6x2ICs}) show that, for this statistic, the large group around $\daa=-1 \times 10^{-5}$ is strongly preferred. This is in stark contrast to the AICc distribution in panel (a) which is far more ambiguous. Interestingly, these two panels also produce quite different values of $\langle \daa \rangle$, as the insets in each panel show.

Whilst the actual values of the AICc and SpIC parameters cannot be easily compared (given their very different definitions), panels (a) and (d) of Figure \ref{fig:6x2nc} illustrate that SpIC requires fewer free parameters than does AICc. This is reflected in the different values of $\langle \sigma_s \rangle$; the AICc mean statistical uncertainty, $\langle \sigma_s \rangle$, is 14\% larger than that of SpIC.

Still referring to HE0515$-$4414, the compound and thermal broadening models seem to produce far more robust results than for turbulent. For compound broadening, the AICc statistic still exhibits two groupings in panels (b) of Figures \ref{fig:6x2ICs}, \ref{fig:6x2nc}, or \ref{fig:6x2chisq}. In contrast, the SpIC distributions in panels (e) are tight and do not indicate multiple minima. This is unsurprising as although compound broadening requires one additional free parameter per velocity component, it is the more general and hence more correct physical model and, interestingly, {\it fewer components are needed to fit the data}. This can be seen, for example, by comparing $\langle n_c \rangle$ in panel (e) of Figure \ref{fig:6x2nc} with panels (d) and (f) and it reinforces the expectation that compound broadening is the appropriate physical model.

The thermal distributions (panels (c) and (f)) show far less empirical scatter than the turbulent case, implying that for this particular absorption system, thermal broadening is closer to reality than turbulent. Whether or not that is generally true is not addressed here but in any case it is suggestive from our results that {\it both} turbulent and thermal models should be avoided and that, compound broadening should be used.

Next we turn to the Q0528$-$2505 spectrum, reminding the reader that the spectral resolution in this case is $R \approx 50,000$ compared to $R \approx 115,000$ for HE0515$-$4414.

\subsection{System 2: $z_{abs}=2.141$ system Q0528$-$2505}

We again begin by examining the turbulent broadening case for the Q0528$-$2505 system. Panels (g) and (j) in any of Figures \ref{fig:6x2ICs}, \ref{fig:6x2nc}, or \ref{fig:6x2chisq} all indicate structure. In panels (g) and (j) of Figure \ref{fig:6x2ICs} for example, both information criteria show there are at least two groupings of points, one around $\daa \approx 0$, the other near $\daa \approx 0.5 \times 10^{-5}$. Model non-uniqueness is thus present for turbulent broadening regardless of whether fitting is carried out using AICc or SpIC. Panels (g) and (j) in Figure \ref{fig:6x2chisq} show a trend between $\daa$ and the normalised $\chi^2$ for the fit. Thus in this case, provided multiple independent fits are obtained, one can in principle identify the higher $\chi^2$ grouping around $\daa \approx 0.5 \times 10^{-5}$ as being spurious. However, it is important to note that non-automated, non-Monte Carlo analysis (i.e. a single interactively-fitted model) could easily fail to identify this property. Notice too that the plotting scale for the Q0528$-$2505 absorption system is far smaller than for HE0515$-$4414, i.e. {\it model non-uniqueness is far less pronounced for the Q0528$-$2505 system compared to the HE0515$-$4414 system}.

Superficially, the compound broadening models illustrated in panels (h) and (k) of Figure \ref{fig:6x2ICs} suggest that AICc models suffer substantially less from non-uniqueness problems than do SpIC models. However, careful inspection shows the following. The empirical scatter for $\daa$ amongst all models in panel (h) is $(0.359 + 0.248) \times 10^{-5} \approx 0.6 \times 10^{-5}$. For panel (k), the SpIC data, it is $(0.305 + 0.381) \times 10^{-5} \approx 0.7 \times 10^{-5}$. On the other hand, the mean statistical uncertainties, $\langle \sigma_s \rangle$, are 0.35 and 0.32 $\times 10^{-5}$ respectively. Thus for both information criteria, the empirical scatter is substantially larger, which indicates non-uniqueness is present in both cases. It so happens for this particular absorption system that segregation of points with respect to the SpIC parameter is more distinct. However, this distinction, for the SpIC sample, is useful because one would justifiably adopt the points below SpIC $\approx 1700$ as being favoured, {\it provided that} a Monte-Carlo sample of models has been calculated.

Interestingly, the same bifurcation seen in panel (k) of Figure \ref{fig:6x2ICs} is not repeated in panel (k) of Figure \ref{fig:6x2nc} although it is in panel (k) of Figure \ref{fig:6x2chisq}. Finally, in Figures \ref{fig:6x2ICs}, \ref{fig:6x2nc}, and \ref{fig:6x2chisq}, panel (k), all exhibit excessive scatter compared to the statistical uncertainty, again caused by non-uniqueness.

\section{Conclusions}

Our detailed study comprises only two quasar absorption systems so of course caution should be exercised in generalising too far. Studies of the sort we have described here should clearly be extended using a larger sample of UVES/VLT, Keck/HIRES, and ESPRESSO/VLT data. Nevertheless, we can already draw some interesting conclusions from HPC calculations we have carried out so far:

\begin{enumerate}[leftmargin=0.5cm]

\item[(1)] The availability of a fully automated and unbiased procedure such as {\sc ai-vpfit} reveals (as expected, but not previously demonstrated) that the $\chi^2$ space may contain multiple local minima. A human computing a single subjective model interactively would most likely select one model at random from the Monte Carlo samples we have illustrated in this paper. This means that it would be difficult or impossible to readily distinguish between a ``real'' and ``fake'' minimum. Multiple independently constructed and unbiased models are needed to reliably interpret the data.

\item[(2)] The effect described above is especially prominent for models derived using turbulent broadening, but increased scatter can also be seen for thermal broadening models. The HPC calculations reported here show that when solving for $\daa$, it is advisable to use the more physically relevant compound line broadening models {\it and} that this is best done in conjunction with the SpIC information criterion rather than AICc. SpIC outperformed AICc on the HE0515$-$4414 analysis. SpIC and AICc performed comparably for the Q0528$-$2505 data. SpIC required fewer free parameters in both cases. Thus overall we suggest SpIC is the preferred information criterion.

\item[(3)] Further to point (2) above, it is interesting to note that when AICc is used in conjunction with turbulent broadening for HE0515$-$4414, non-uniqueness is severe, yet when SpIC is used in conjunction with compound broadening, there is no indication of non-uniqueness whatsoever. The same phenomenon is not seen in the lower resolution data for Q0528$-$2505. Whilst we therefore cannot confidently generalise, it is very likely that the combination of SpIC plus compound broadening will yield the most reliable $\daa$ measurements.

\item[(4)] Interactive measurements, in which a single model is calculated, suffer from non-uniqueness, imposing an additional uncertainty that is substantially larger than the usual statistical uncertainty. This means that estimating $\daa$ is, most probably, inherently a {\it statistical} problem i.e. ultimately we will only be able to demonstrate the existence (or non-existence) of spacetime variation of fundamental constants using statistical samples of measurements; for a reliable interpretation of high resolution quasar spectra, measurement of $\daa$ should involve estimating the non-uniqueness ambiguity, on a system-by-system basis. Of course, if an absorption system is found whose kinematics just happen to be such that little or no non-uniqueness is present, that system may indeed be a ``holy grail'' and yield a very precise measurement of $\daa$.

\end{enumerate}

\section*{Acknowledgements}
We are grateful for supercomputer time using OzSTAR at the Centre for Astrophysics and Supercomputing at Swinburne University of Technology. CCL thanks the Royal Society for a Newton International Fellowship during the early stages of this work. JKW thanks the John Templeton Foundation, the Department of Applied Mathematics and Theoretical Physics, the Institute of Astronomy, and Clare Hall at Cambridge University for hospitality and support. 

\section*{Data Availability}
The observational data were collected at the European Southern Observatory under ESO programme 102.A-0697(A) and are available through the ESO archive. The {\sc ai-vpfit} models can be obtained from the authors on request.

\bibliographystyle{mnras}
\bibliography{nonuniqueness}

\begin{thebibliography}{}
\makeatletter
\relax
\def\mn@urlcharsother{\let\do\@makeother \do\$\do\&\do\#\do\^\do\_\do\%\do\~}
\def\mn@doi{\begingroup\mn@urlcharsother \@ifnextchar [ {\mn@doi@}
  {\mn@doi@[]}}
\def\mn@doi@[#1]#2{\def\@tempa{#1}\ifx\@tempa\@empty \href
  {http://dx.doi.org/#2} {doi:#2}\else \href {http://dx.doi.org/#2} {#1}\fi
  \endgroup}
\def\mn@eprint#1#2{\mn@eprint@#1:#2::\@nil}
\def\mn@eprint@arXiv#1{\href {http://arxiv.org/abs/#1} {{\tt arXiv:#1}}}
\def\mn@eprint@dblp#1{\href {http://dblp.uni-trier.de/rec/bibtex/#1.xml}
  {dblp:#1}}
\def\mn@eprint@#1:#2:#3:#4\@nil{\def\@tempa {#1}\def\@tempb {#2}\def\@tempc
  {#3}\ifx \@tempc \@empty \let \@tempc \@tempb \let \@tempb \@tempa \fi \ifx
  \@tempb \@empty \def\@tempb {arXiv}\fi \@ifundefined
  {mn@eprint@\@tempb}{\@tempb:\@tempc}{\expandafter \expandafter \csname
  mn@eprint@\@tempb\endcsname \expandafter{\@tempc}}}

\bibitem[\protect\citeauthoryear{{Akaike}}{{Akaike}}{1974}]{Akaike1974}
{Akaike} H.,  1974, IEEE Transactions on Automatic Control, \href
  {http://adsabs.harvard.edu/abs/1974ITAC...19..716A} {19, 716}

\bibitem[\protect\citeauthoryear{{Bainbridge} \& {Webb}}{{Bainbridge} \&
  {Webb}}{2017a}]{Bainbridge2017}
{Bainbridge} M.~B.,  {Webb} J.~K.,  2017a, \mn@doi [Universe]
  {10.3390/universe3020034}, \href
  {https://ui.adsabs.harvard.edu/#abs/2017Univ....3...34B} {3, 34}

\bibitem[\protect\citeauthoryear{{Bainbridge} \& {Webb}}{{Bainbridge} \&
  {Webb}}{2017b}]{gvpfit17}
{Bainbridge} M.~B.,  {Webb} J.~K.,  2017b, \mn@doi [MNRAS]
  {10.1093/mnras/stx179}, \href
  {http://adsabs.harvard.edu/abs/2017MNRAS.468.1639B} {468, 1639}

\bibitem[\protect\citeauthoryear{{Barrow} \& {Lip}}{{Barrow} \&
  {Lip}}{2012}]{Barrow2012}
{Barrow} J.~D.,  {Lip} S. Z.~W.,  2012, \mn@doi [Phys. Rev.~D]
  {10.1103/PhysRevD.85.023514}, \href
  {https://ui.adsabs.harvard.edu/#abs/2012PhRvD..85b3514B} {85, 023514}

\bibitem[\protect\citeauthoryear{{Bozdogan}}{{Bozdogan}}{1987}]{Bozdogan1987}
{Bozdogan} H.,  1987, \mn@doi [Psychometrika] {10.1007/BF02294361}, 52, 345

\bibitem[\protect\citeauthoryear{{Carswell} \& {Webb}}{{Carswell} \&
  {Webb}}{2014}]{VPFIT}
{Carswell} R.~F.,  {Webb} J.~K.,  2014, {VPFIT: Voigt profile fitting program},
  Astrophysics Source Code Library (\mn@eprint {ascl} {1408.015})

\bibitem[\protect\citeauthoryear{{Carswell} \& {Webb}}{{Carswell} \&
  {Webb}}{2020}]{web:VPFIT}
{Carswell} R.~F.,  {Webb} J.~K.,  2020, Bob Carswell's homepage, \url
  {https://people.ast.cam.ac.uk/~rfc/}

\bibitem[\protect\citeauthoryear{{Hurvich} \& {Tsai}}{{Hurvich} \&
  {Tsai}}{1989}]{Hurvich1989}
{Hurvich} C.~M.,  {Tsai} C.-L.,  1989, \mn@doi [Biometrika]
  {10.1093/biomet/76.2.297}, 76, 297

\bibitem[\protect\citeauthoryear{{King}, {Mortlock}, {Webb}  \&
  {Murphy}}{{King} et~al.}{2009}]{King2009}
{King} J.~A.,  {Mortlock} D.~J.,  {Webb} J.~K.,   {Murphy} M.~T.,  2009,
  \memsai, \href {https://ui.adsabs.harvard.edu/abs/2009MmSAI..80..864K} {80,
  864}

\bibitem[\protect\citeauthoryear{{King}, {Webb}, {Murphy}, {Flambaum},
  {Carswell}, {Bainbridge}, {Wilczynska}  \& {Koch}}{{King}
  et~al.}{2012}]{King2012}
{King} J.~A.,  {Webb} J.~K.,  {Murphy} M.~T.,  {Flambaum} V.~V.,  {Carswell}
  R.~F.,  {Bainbridge} M.~B.,  {Wilczynska} M.~R.,   {Koch} F.~E.,  2012,
  \mn@doi [MNRAS] {10.1111/j.1365-2966.2012.20852.x}, \href
  {https://ui.adsabs.harvard.edu/abs/2012MNRAS.422.3370K} {422, 3370}

\bibitem[\protect\citeauthoryear{{Kotu{\v{s}}}, {Murphy}  \&
  {Carswell}}{{Kotu{\v{s}}} et~al.}{2017}]{Kotus2017}
{Kotu{\v{s}}} S.~M.,  {Murphy} M.~T.,   {Carswell} R.~F.,  2017, \mn@doi
  [MNRAS] {10.1093/mnras/stw2543}, \href
  {https://ui.adsabs.harvard.edu/abs/2017MNRAS.464.3679K} {464, 3679}

\bibitem[\protect\citeauthoryear{{Lee}, {Webb}, {Carswell}  \&
  {Milakovi{\'c}}}{{Lee} et~al.}{2021}]{Lee2020AI-VPFIT}
{Lee} C.-C.,  {Webb} J.~K.,  {Carswell} R.~F.,   {Milakovi{\'c}} D.,  2021,
  \mn@doi [\mnras] {10.1093/mnras/stab977}, \href
  {https://ui.adsabs.harvard.edu/abs/2021MNRAS.504.1787L} {504, 1787}

\bibitem[\protect\citeauthoryear{Marconi et~al.,}{Marconi
  et~al.}{2016}]{Marconi2016}
Marconi A.,  et~al., 2016, in {Ground-based and Airborne Instrumentation for
  Astronomy VI, eds. Christopher J. Evans and Luc Simard and Hideki Takami}.
  SPIE, pp 676 -- 687, \mn@doi{10.1117/12.2231653}, \url
  {https://doi.org/10.1117/12.2231653}

\bibitem[\protect\citeauthoryear{{Milakovi{\'c}}, {Pasquini}, {Webb}  \& {Lo
  Curto}}{{Milakovi{\'c}} et~al.}{2020}]{Milakovic2020}
{Milakovi{\'c}} D.,  {Pasquini} L.,  {Webb} J.~K.,   {Lo Curto} G.,  2020,
  \mn@doi [\mnras] {10.1093/mnras/staa356}, \href
  {https://ui.adsabs.harvard.edu/abs/2020MNRAS.493.3997M} {493, 3997}

\bibitem[\protect\citeauthoryear{{Milakovi{\'c}}, {Lee}, {Carswell}, {Webb},
  {Molaro}  \& {Pasquini}}{{Milakovi{\'c}} et~al.}{2021}]{Milakovic2021}
{Milakovi{\'c}} D.,  {Lee} C.-C.,  {Carswell} R.~F.,  {Webb} J.~K.,  {Molaro}
  P.,   {Pasquini} L.,  2021, \mn@doi [\mnras] {10.1093/mnras/staa3217}, \href
  {https://ui.adsabs.harvard.edu/abs/2021MNRAS.500....1M} {500, 1}

\bibitem[\protect\citeauthoryear{{Murphy}, {Kacprzak}, {Savorgnan}  \&
  {Carswell}}{{Murphy} et~al.}{2019}]{Murphy2019}
{Murphy} M.~T.,  {Kacprzak} G.~G.,  {Savorgnan} G. A.~D.,   {Carswell} R.~F.,
  2019, \mn@doi [\mnras] {10.1093/mnras/sty2834}, \href
  {https://ui.adsabs.harvard.edu/abs/2019MNRAS.482.3458M} {482, 3458}

\bibitem[\protect\citeauthoryear{{Pepe} et~al.,}{{Pepe}
  et~al.}{2021}]{espresso2021}
{Pepe} F.,  et~al., 2021, \mn@doi [\aap] {10.1051/0004-6361/202038306}, \href
  {https://ui.adsabs.harvard.edu/abs/2021A&A...645A..96P} {645, A96}

\bibitem[\protect\citeauthoryear{{Probst} et~al.,}{{Probst}
  et~al.}{2020}]{Probst2020}
{Probst} R.~A.,  et~al., 2020, \mn@doi [Nature Astronomy]
  {10.1038/s41550-020-1010-x}, \href
  {https://ui.adsabs.harvard.edu/abs/2020NatAs...4..603P} {4, 603}

\bibitem[\protect\citeauthoryear{{Tamai}, {Koehler}, {Cirasuolo},
  {Biancat-Marchet}, {Tuti}  \& {Gonz{\'a}les Herrera}}{{Tamai}
  et~al.}{2018}]{ELT2018}
{Tamai} R.,  {Koehler} B.,  {Cirasuolo} M.,  {Biancat-Marchet} F.,  {Tuti} M.,
   {Gonz{\'a}les Herrera} J.~C.,  2018, in {Marshall} H.~K.,  {Spyromilio} J.,
  eds,  Society of Photo-Optical Instrumentation Engineers (SPIE) Conference
  Series Vol. 10700, Ground-based and Airborne Telescopes VII. p. 1070014,
  \mn@doi{10.1117/12.2309515}

\bibitem[\protect\citeauthoryear{{Webb}, {King}, {Murphy}, {Flambaum},
  {Carswell}  \& {Bainbridge}}{{Webb} et~al.}{2011}]{Webb2011}
{Webb} J.~K.,  {King} J.~A.,  {Murphy} M.~T.,  {Flambaum} V.~V.,  {Carswell}
  R.~F.,   {Bainbridge} M.~B.,  2011, \mn@doi [Phys. Rev.~Lett.]
  {10.1103/PhysRevLett.107.191101}, \href
  {https://ui.adsabs.harvard.edu/abs/2011PhRvL.107s1101W} {107, 191101}

\bibitem[\protect\citeauthoryear{{Webb}, {Lee}, {Carswell}  \&
  {Milakovi{\'c}}}{{Webb} et~al.}{2021}]{Webb2021}
{Webb} J.~K.,  {Lee} C.-C.,  {Carswell} R.~F.,   {Milakovi{\'c}} D.,  2021,
  \mn@doi [\mnras] {10.1093/mnras/staa3551}, \href
  {https://ui.adsabs.harvard.edu/abs/2021MNRAS.501.2268W} {501, 2268}

\makeatother
\end{thebibliography}

\appendix

\section{{\sc ai-vpfit} AICc models with compound broadening for HE0515$-$4414} \label{sec:AICcmodels}

\begin{figure*}
\centering
\includegraphics[width=0.9\linewidth]{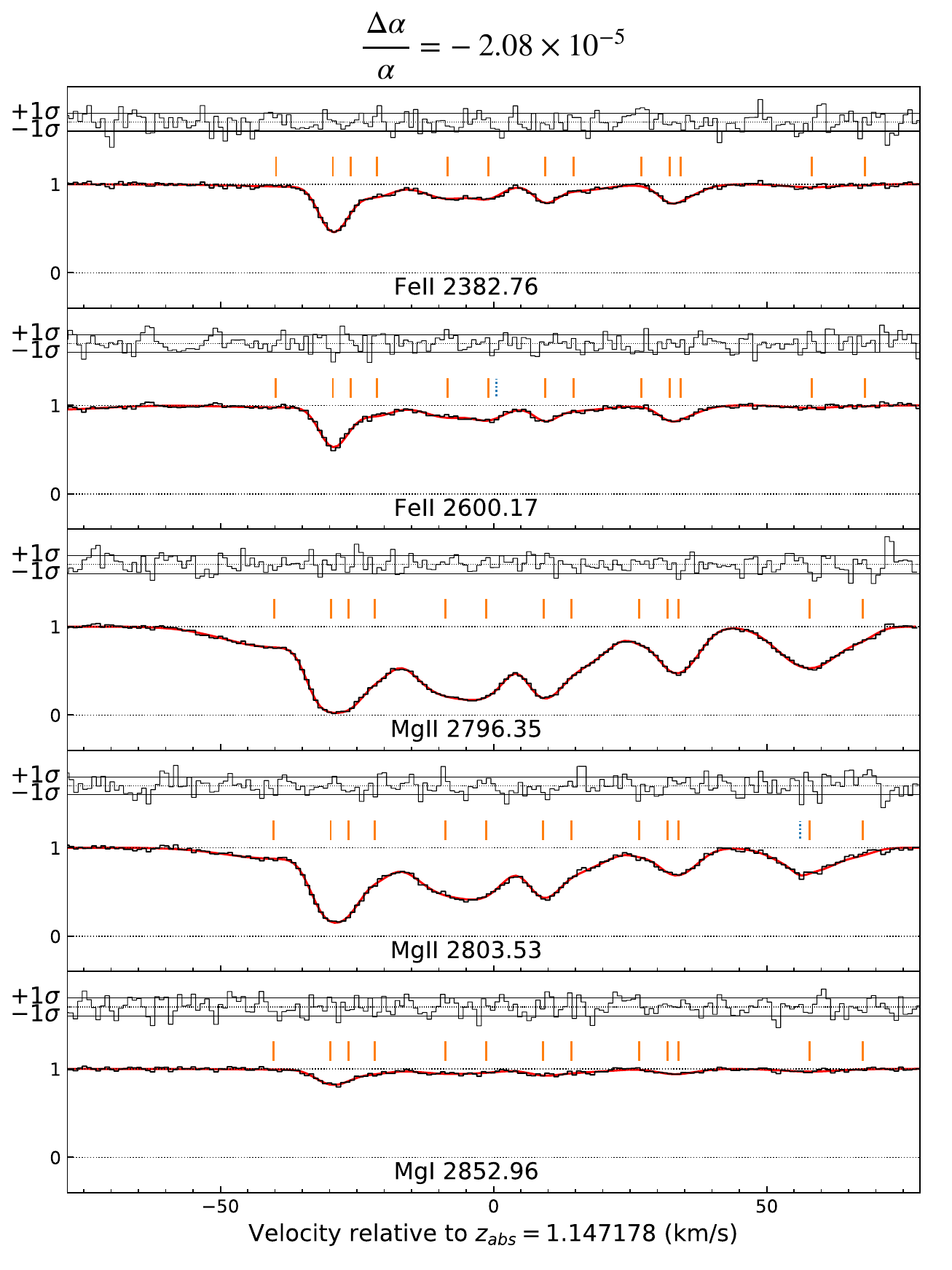}
\caption{ Demonstration of the non-uniqueness problem. The black histogram shows the $z_{abs}=1.147$ towards the quasar HE0515$-$4414, The atomic transition is indicated in each case. The best fit model is obtained using the AICc and compound line broadening. For this particular model, $\daa = -2.08 \times 10^{-5}$.}
\label{fig:spec_HE0515a}
\end{figure*}

\begin{figure*}
\centering
\includegraphics[width=0.9\linewidth]{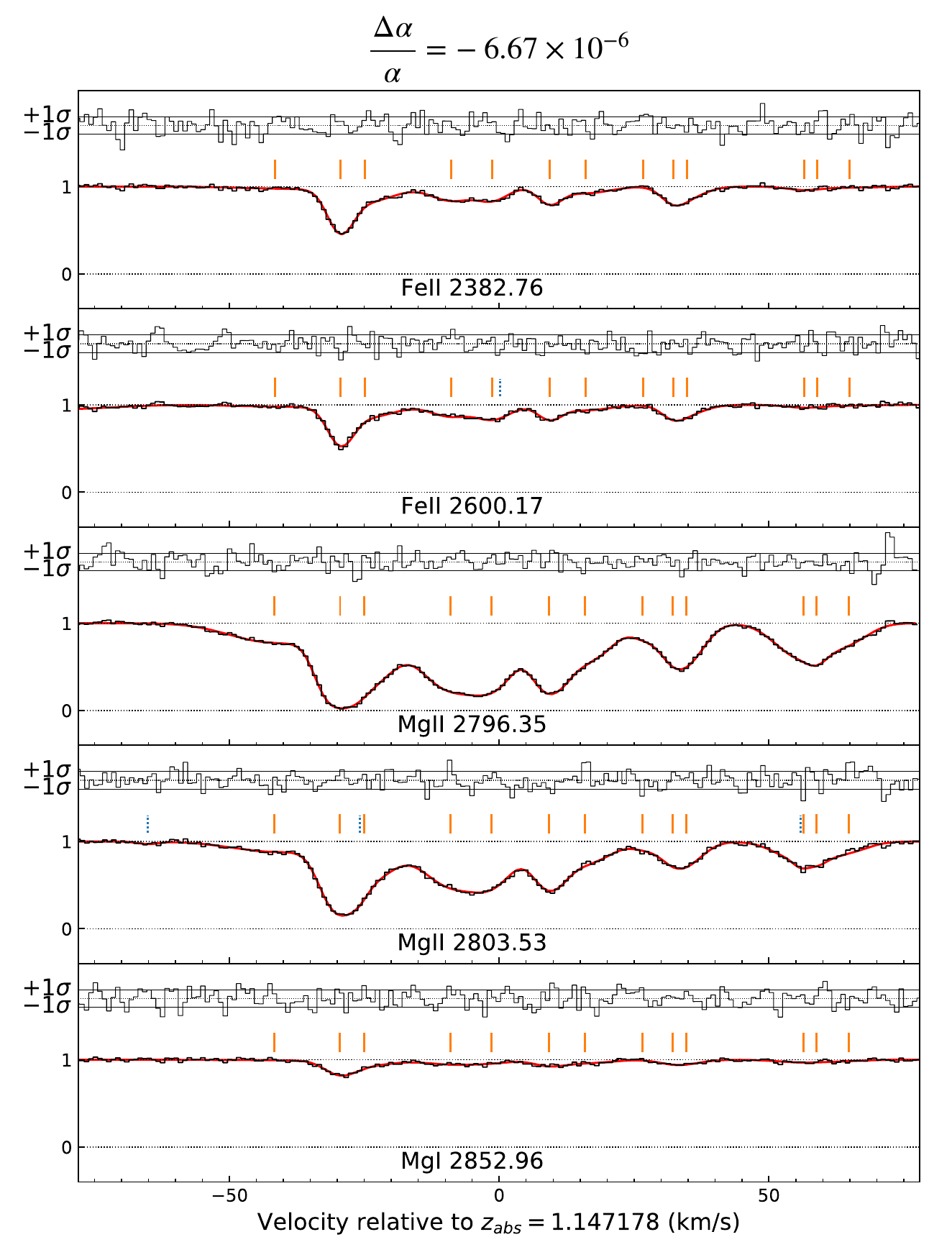}
\caption{ Same as Fig.~\ref{fig:spec_HE0515a} except AI-VPFIT model construction used a different set of random seeds. Model development is unique to each set of random seeds. For this particular model, $\daa = -6.67 \times 10^{-6}$.}
\label{fig:spec_HE0515b}
\end{figure*}

\begin{figure*}
\centering
\includegraphics[width=0.9\linewidth]{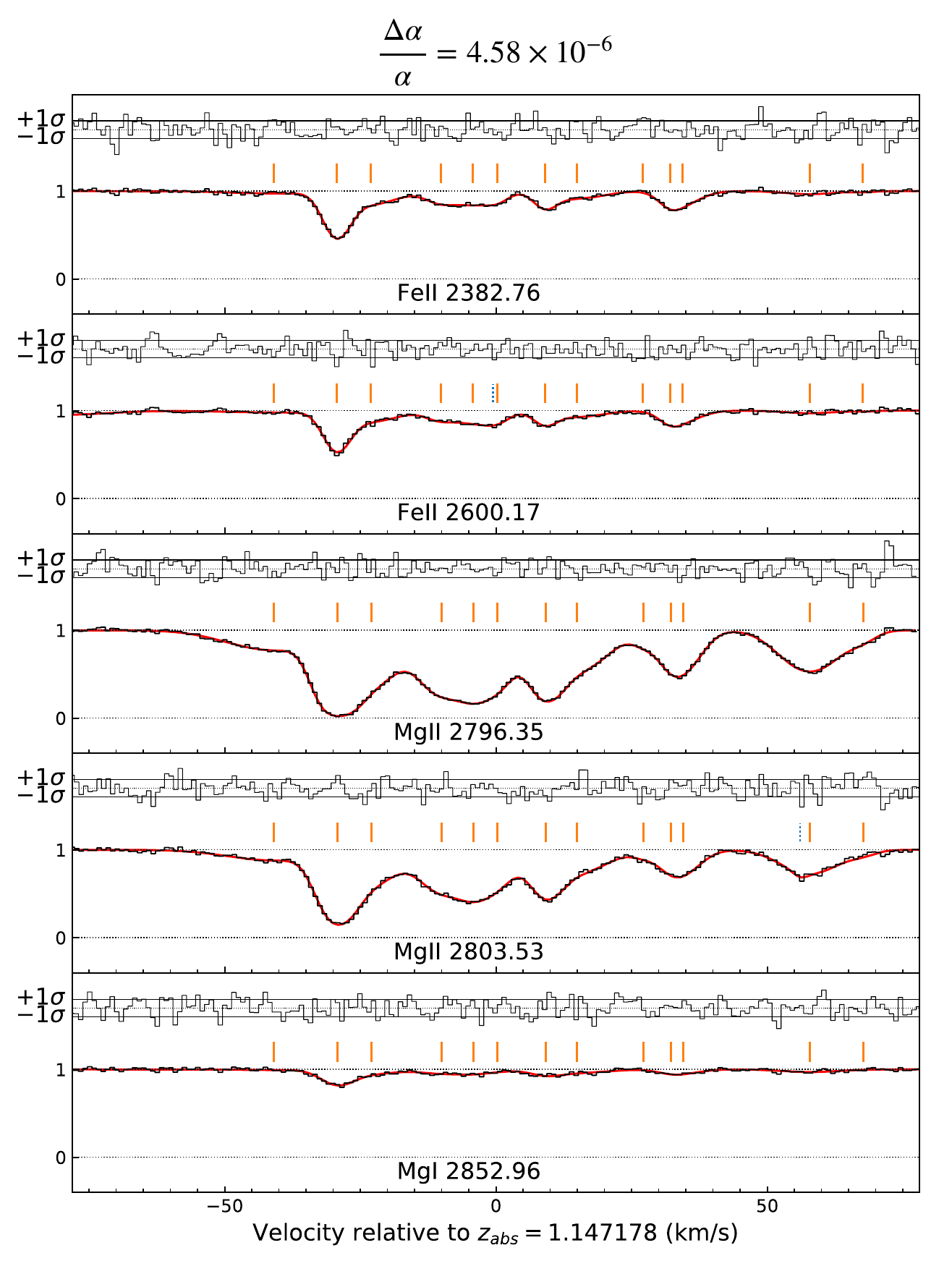}
\caption{ Same as Fig.~\ref{fig:spec_HE0515a}, with a new set of random seeds. For this particular model, $\daa = 4.58 \times 10^{-6}$.}
\label{fig:spec_HE0515c}
\end{figure*}

\begin{figure*}
\centering
\includegraphics[width=0.9\linewidth]{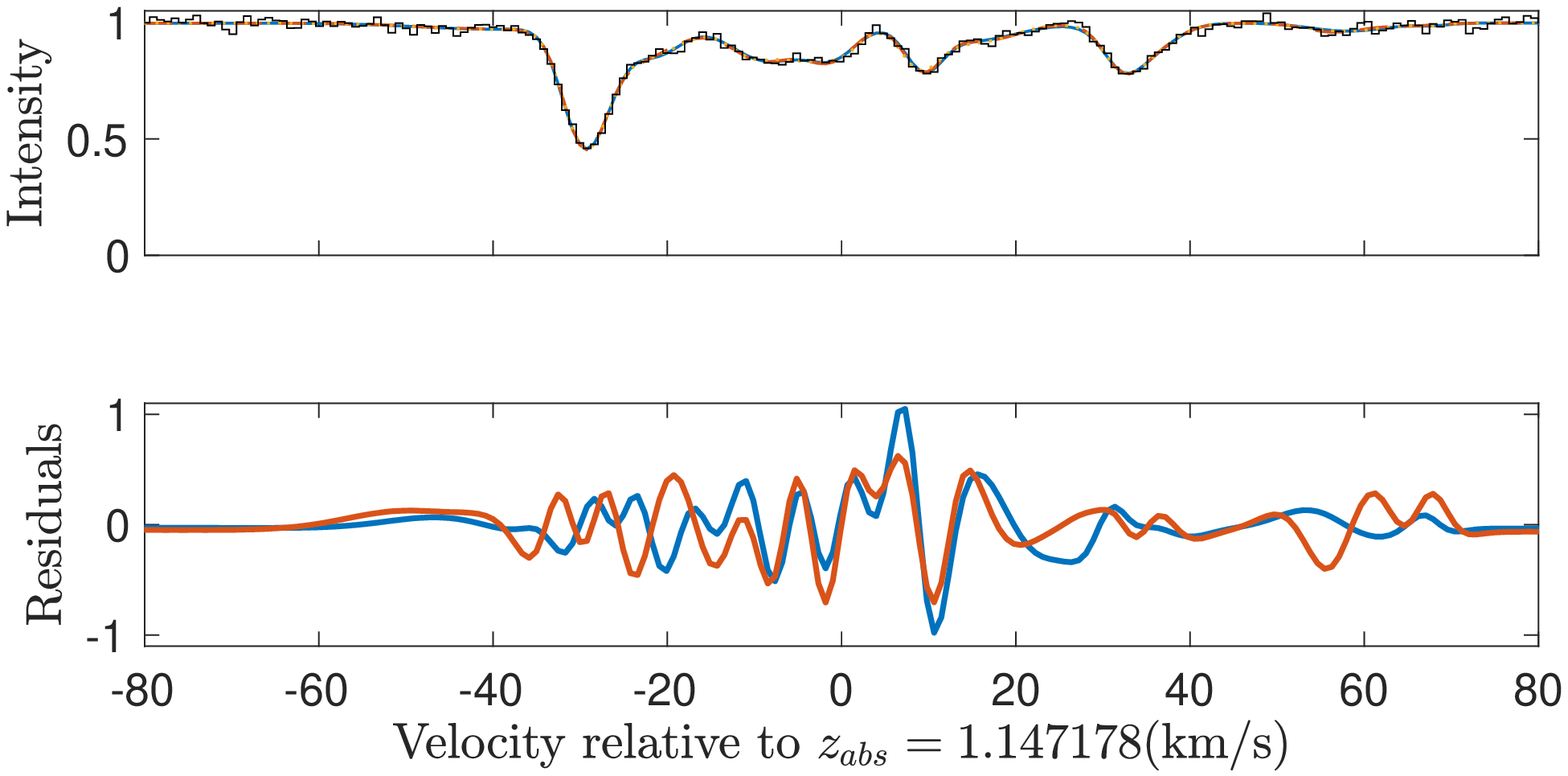}
\caption{The top panel illustrates one transition (Fe\,{\sc ii} 2383) from the $z_{abs}=1.147$ towards the quasar HE0515$-$4414. The black histogram shows the observed spectrum. The blue continuous line shows model A1, the red dashed line shows model A2, and the yellow dotted line shows model A3. The bottom panel gives the difference between the models; the blue continuous line shows A1-A3 and the red continuous line shows A2-A3. In each case the difference is divided by the spectral error array i.e. the y-axis is a normalised residual. The differences fluctuate on scales reaching to around $\pm 1 \sigma$.}
\label{fig:spec_diff}
\end{figure*}

The plots illustrated in Figures \ref{fig:spec_HE0515a}, \ref{fig:spec_HE0515b}, \ref{fig:spec_HE0515c}, and \ref{fig:spec_diff} relate to HE0515$-$4414. The line broadening mechanism is compound but the model selection criterion is AICc. Comparing panels (b) with panels (e) in Figures \ref{fig:6x2ICs}, \ref{fig:6x2nc}, and \ref{fig:6x2chisq} show that {\sc ai-vpfit} models derived using AICc tend to be overfitted, causing AICc to generate unnecessary model ambiguity. This problem is avoided when SpIC is used. Here we illustrate three AICc models spanning the range in $\daa$ shown in panels (b).

\clearpage
\section{The Q0528$-$2505 $z_{abs}=2.141$ model}

\begin{figure*}
\centering
\includegraphics[width=0.9\linewidth]{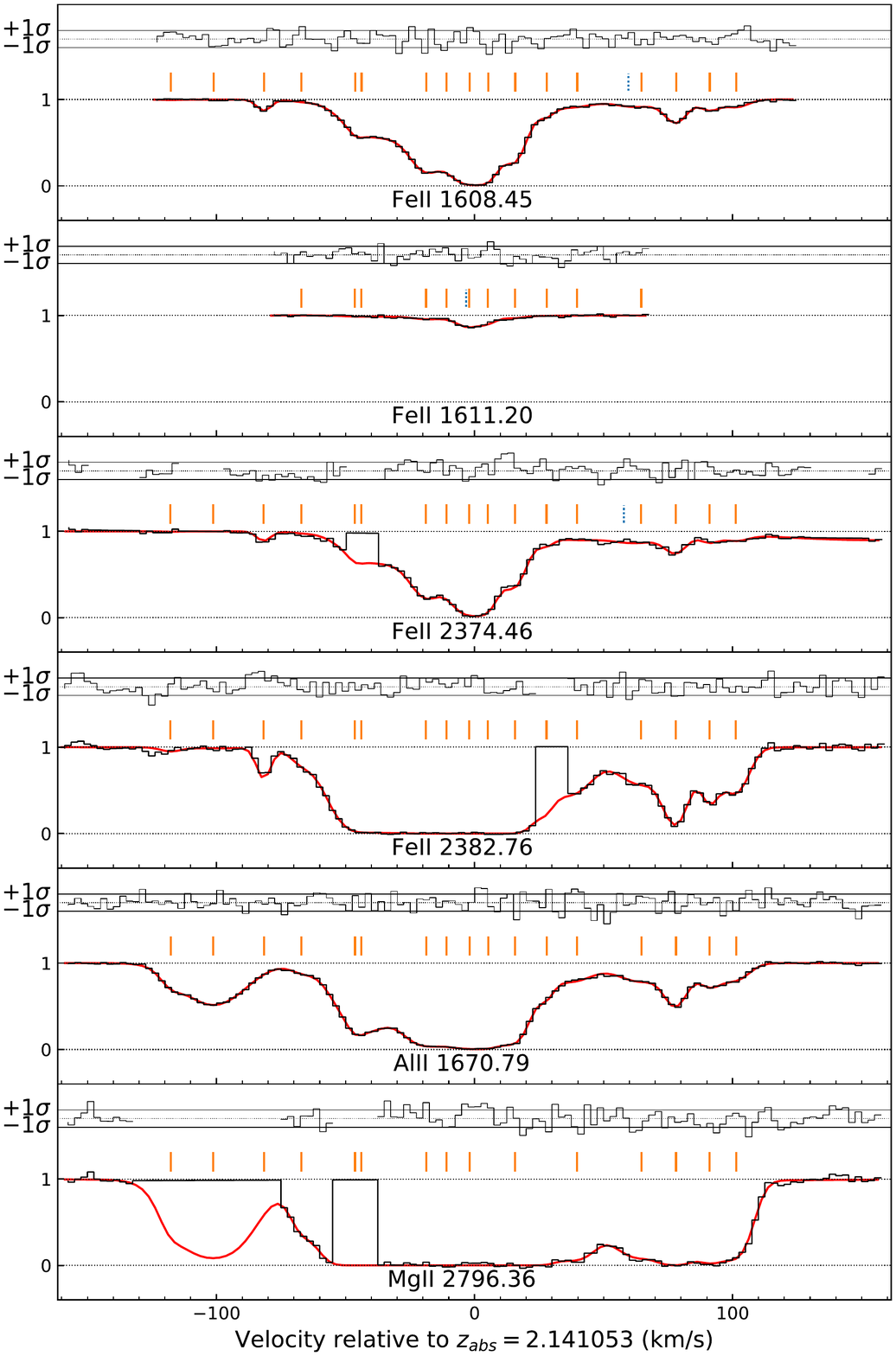}
\caption{This figure corresponds to Figure \ref{fig:spec_HE0515a} except this is now for the $z_{abs}=2.141$ system towards the quasar Q0528$-$2505. This particular model gives $\daa = 3.46\times 10^{-6}$. Three transitions (Fe\,{\sc ii} 2374, 2382, and Mg\,{\sc ii} 2796, have small regions missing due to cosmic ray events on the CCD. These regions have been excluded from the fit.}
\label{fig:Q0528spec_a}
\end{figure*}

\begin{figure*}
\centering
\includegraphics[width=0.9\linewidth]{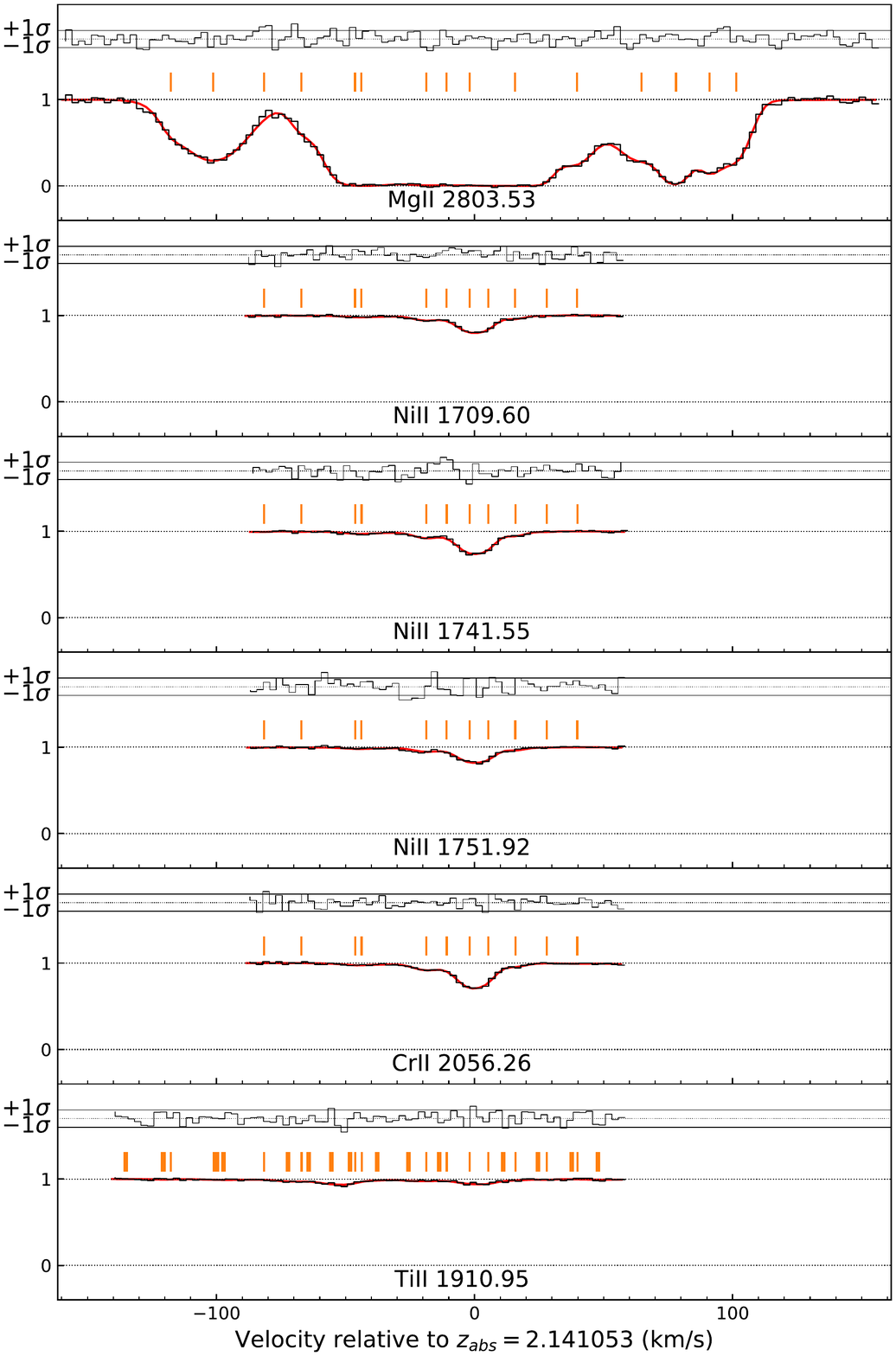}
\caption{Continuation of Figure \ref{fig:Q0528spec_a}. The extended set of orange tick marks for Ti\,{\sc ii} 1910 arises from a close blend of two lines (1910.9538 and 1910.6123{\AA}.)}
\label{fig:Q0528spec_b}
\end{figure*}

\bsp
\label{lastpage}
\end{document}